\documentclass[prb,twocolumn,showpacs,amsmath,amssymb,superscriptaddress]{revtex4}
\usepackage[dvips]{graphicx}
\usepackage{graphics}
\usepackage{dcolumn}
\usepackage{bm}

\begin{document}

\title{Element-resolved orbital polarization in (III,Mn)As ferromagnetic semiconductors from $K$ edge x-ray magnetic circular dichroism} 

\author{P.~Wadley}
\affiliation{School of Physics and Astronomy, University of Nottingham, Nottingham NG7 2RD, United Kingdom}
\author{A.~A.~Freeman}
\affiliation{School of Physics and Astronomy, University of Nottingham, Nottingham NG7 2RD, United Kingdom}
\affiliation{Diamond Light Source, Chilton, Didcot OX11 0DE, United Kingdom}
\author{K.~W.~Edmonds}
\affiliation{School of Physics and Astronomy, University of Nottingham, Nottingham NG7 2RD, United Kingdom}
\author{G.~van der Laan}
\affiliation{Diamond Light Source, Chilton, Didcot OX11 0DE, United Kingdom}
\author{J.~S.~Chauhan}
\author{R.~P.~Campion}
\author{A.~W.~Rushforth}
\author{B.~L.~Gallagher}
\author{C.~T.~Foxon}
\affiliation{School of Physics and Astronomy, University of Nottingham, Nottingham NG7 2RD, United Kingdom}
\author{F.~Wilhelm}
\author{A.~G.~Smekhova}
\author{A.~Rogalev}
\affiliation{European Synchrotron Radiation Facility, F-38043 Grenoble Cedex 9, France}

\date{\today}
\begin{abstract}
Using x-ray magnetic circular dichroism (XMCD), we determine the element-specific character and polarization of unoccupied states near the Fermi level in (Ga,Mn)As and (In,Ga,Mn)As thin films. The XMCD at the As $K$ absorption edge consists of a single peak located on the low-energy side of the edge, which increases with the concentration of ferromagnetic Mn moments. The XMCD at the Mn $K$ edge is more detailed and is strongly concentration-dependent, which is interpreted as a signature of hole localization for low Mn doping. The results indicate a markedly different character of the polarized holes in low-doped insulating and high-doped metallic films, with a transfer of the hole orbital magnetic moment from Mn to As sites on crossing the metal-insulator transition.
\end{abstract}

\maketitle

\section{Introduction}

Mn-doped III-V semiconductors are valuable materials for investigating the interplay of itinerant and localized magnetism, as demonstrated in a range of spin-based electronic device structures. \cite{ohno00,sawicki10} The substitutional Mn dopant in Ga$_{1-x}$Mn$_x$As has a large 3$d$ magnetic moment and is also an acceptor, providing holes that mediate ferromagnetic ordering. The magnetic and electrical properties of these materials are thus closely related, and also highly tunable through electrical gating or co-doping. Beyond a critical Mn concentration $x$ of around 2\%, Ga$_{1-x}$Mn$_x$As undergoes a transition from an insulating to a metallic conductivity. \cite{matsukura98} In the highly doped, metallic regime, the interrelated magnetic and electrical properties, including the dependence of the ferromagnetic Curie temperature $T_{\mathrm{C}}$ on the hole density, are in reasonable agreement with calculations based on a model of holes occupying the exchange-split host-like valence band. \cite{sawicki10,dietl01,jungwirth05,rushforth07} In contrast, models involving hopping within a narrow impurity band are used to describe conduction and ferromagnetism in the insulating regime. \cite{kaminski02,kennett02,sheu07} Elsewhere, infrared spectroscopy and electrical transport measurements have been interpreted within a model of holes residing in a narrow impurity band, detached from the valence band, even for highly doped films. \cite{burch06,mayer10} The accuracy of these competing models remains a topic of considerable debate.

Part of the difficulty in characterizing these materials lies with the fact that the concentration of substitutional Mn needed for ferromagnetism is well above the equilibrium solubility limit. Therefore, during growth a high concentration of compensating defects may be incorporated, including antisite As and interstitial Mn, which effectively reduce the concentration of Mn that participates in the ferromagnetic order. \cite{korzhavyi02,yu02} Annealing at temperatures $\sim$200$^\circ$C promotes the diffusion and eventual removal from the lattice of interstitial Mn. \cite{edmonds04} Therefore, the $T_{\mathrm{C}}$, conductivity and density of magnetic moments are found to be higher in annealed films than in unannealed films. \cite{yu02,edmonds04,potashnik02} 

Magnetic circular dichroism techniques have been widely used as probes of electronic structure in ferromagnetic semiconductors. These methods, based on the difference in absorption of left- and right-circularly polarized photons, offer advantages over unpolarized spectroscopy techniques in that they directly probe states that are participating in the magnetic order. The majority of studies have been performed at infrared or optical wavelengths, where the photons excite transitions between valence, impurity and conduction bands. \cite{ando08,berciu09,turek09,acbas09} While these techniques are sensitive to the character of the states around the Fermi energy $E_{\mathrm{F}}$, the complicated nature of the inter- and intra-band transitions involved means that results are not unambiguous; agreement with both impurity band and disordered valence band models of ferromagnetism have been reported. \cite{ando08,berciu09,turek09,acbas09}

X-ray magnetic circular dichroism (XMCD) involves the excitation of core level electrons to unoccupied states in the vicinity of $E_{\mathrm{F}}$, and thus offers an \emph{element-specific} probe of the polarized valence states. The technique has been utilized to reveal details of both Mn and host ions' contributions to magnetic ordering in (Ga,Mn)As, \cite{keavney03,edmonds06,freeman08} as well as in other members of the (III,Mn)V family. \cite{sarigiannidou06,stone06,freeman07} By utilizing sum rules, \cite{thole92,carra93} the technique allows the quantitative and element-specific determination of orbital and spin magnetic moments per ion within the probed layer.

Here, we utilize XMCD at the As and Mn $K$ absorption edges, in which the absorption of x-rays by the 1$s$ core level electrons results in transitions to unoccupied valence states, to investigate the element-specific properties of a series of (Ga,Mn)As as well as (In,Ga,Mn)As films. The XMCD at the As $K$ edge is sensitive to the orbital polarization of valence states with As 4$p$ character, while at the Mn $K$ edge it is sensitive to hybridized Mn 4$p$ states. We substantially extend our previous study \cite{freeman08} by revealing qualitative and quantitative trends in the spectra on increasing the Mn concentration and on low-temperature annealing. We observe striking changes on increasing the Mn concentration beyond the metal-insulator transition: the Mn $K$ XMCD decreases dramatically, accompanied by the emergence of a significant As 4$p$ polarization. 

\section{Sample and experiment details}

The Ga$_{1-x}$Mn$_x$As and (In$_y$Ga$_{1-y}$)$_{1-x}$Mn$_x$As films were grown by molecular beam epitaxy on GaAs(001) and InP(001) substrates respectively, using methods described elsewhere. \cite{freeman08,campion03,wang08} The nominal Mn concentrations were estimated from the Mn/Ga flux ratio; since some of the Mn ions are expected to be incorporated in interstitial or random positions, \cite{yu02} the substitutional Mn concentrations are expected to be smaller than the nominal values. Some of the films were annealed in air at 190$^\circ$C for several hours, which is an established procedure for increasing the density of carriers and uncompensated local moments, due to out-diffusion of compensating interstitial defects. \cite{yu02,edmonds04,potashnik02} The samples were characterized using x-ray diffraction, superconducting quantum interference device (SQUID) magnetometry, and electrical transport measurements. The properties of the studied samples are summarized in Table~\ref{tab:1}. 

\begin{table}   
\begin{tabular}{lccclcc}
\hline \hline
Sample & $d$ & $x$ &  $y$ & Annealed & $T_{\mathrm{C}}$  & $M_S$\\
 & (nm) & \% &  \% & & (K) & (emu/cm$^3$) \\
\hline
G1 & 1500 & 1 & 0 & No & 11 & 5\\
G2 & 1000 & 2 & 0 & No & 30 & 9\\
G3 & 1000 & 5 & 0 & No & 50 & 26\\
G4 & 200 & 5 & 0 & Yes & 101 & 37\\ 
I1 & 500 & 8 & 50 & No & 27 & 18\\
I2 & 200 & 8 & 40 & Yes & 62 & 29\\
I3 & 140 & 8 & 60 & Yes & 68 & 35\\
\hline \hline
\end{tabular}
\caption{\label{tab:1} Properties of the studied samples: thickness ($d$) of Mn-doped layer, nominal Mn ($x$) and In ($y$) concentrations estimated from flux ratios, Curie temperature ($T_{\mathrm{C}}$) and saturation magnetization ($M_S$) measured by SQUID magnetometry.}
\label{table1}
\end{table}

\begin{figure}[h]    
\centering
\includegraphics[width=3.1in]{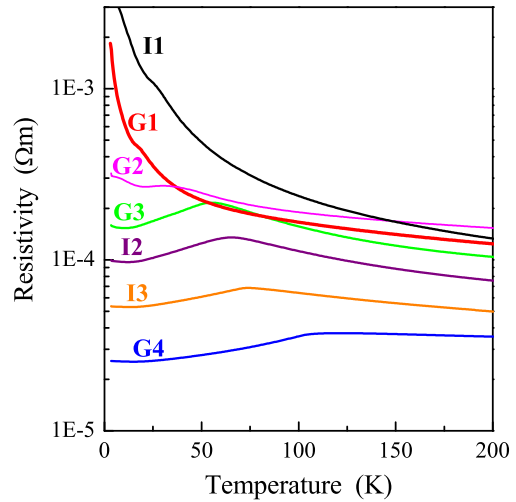}
\caption{\label{fig:1} (Color online)
Resistivity versus temperature curves for the (Ga,Mn)As (G1-G4) and (In,Ga,Mn)As (I1-I3) samples used in this study.}
\end{figure}

Figure~\ref{fig:1} shows resistivity $\rho$ versus temperature traces for the studied films. Consistent with previous studies, \cite{matsukura98} the films are insulating in the low Mn doped (G1, G2) or highly compensated (I1) samples and metallic in the more highly doped and annealed samples (G4, I2, I3). A clear correlation between conductivity and $T_{\mathrm{C}}$ is evident, with the samples having higher $T_{\mathrm{C}}$ exhibiting higher conductivity and more metallic behavior. The $T_{\mathrm{C}}$ coincides approximately with a peak or a shoulder in the $\rho$ vs. $T$ curves in the metallic or insulating samples, respectively, and can be accurately identified from the position of a peak in the first derivative $d\rho/dT$. \cite{novak08}

The attenuation depth of x-rays around the As $K$ edge is several microns. Therefore, in order to prevent the substrate from dominating the absorption signal, the (Ga,Mn)As films were attached face-down onto an Al$_2$O$_3$(0001) wafer using photoresist, and the GaAs substrate was removed by etching away a 100-200~nm thick AlAs buffer layer using concentrated HF. SQUID measurements were performed on the etched films to ensure that this procedure does not significantly affect the $T_{\mathrm{C}}$ of the film. For the (In,Ga,Mn)As films, the measured absorption and XMCD signals were corrected to account for the presence of a thin (25-50~nm) buffer layer between the film and the InP substrate. 

The x-ray absorption and XMCD measurements were performed using 98\% circularly polarized x-rays on beamline ID12 of the European Synchrotron Radiation Facility. Mn $K$ edge measurements were performed on samples G1-G4 and I1, and As $K$ edge measurements were performed on samples G2-G4 and I1-I3. All of the absorption spectra were recorded in total fluorescence yield mode, for either normal or grazing (15$^\circ$) incidence of the x-ray beam. The XMCD signal was obtained from the difference in absorption for parallel and antiparallel alignments of the x-ray polarization vector with respect to an external magnetic field of up to 6~T, applied along the beam direction. Measurements, taken over several hours, were averaged in order to improve the signal-to-noise ratio. To avoid experimental artifacts, the x-ray helicity and the external magnetic field direction were alternately flipped. The XMCD signals are given as a percentage of the absorption edge jump.

\section{Results and discussion}

\subsection{As $K$ edge XMCD}

\begin{figure}[h]    
\centering
\includegraphics[width=3.1in]{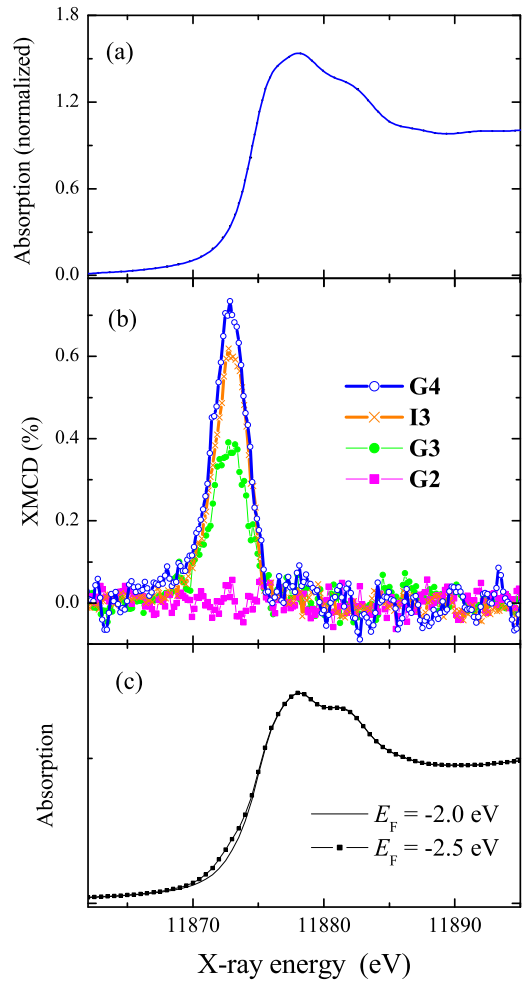}
\caption{\label{fig:2} (Color online)
(a) As $K$ edge x-ray absorption spectrum; (b) XMCD spectra for (Ga,Mn)As samples G2 (2\% Mn, as-grown, squares), G3 (5\% Mn, as-grown, filled circles), G4 (5\% Mn, annealed, open circles), and (In,Ga,Mn)As sample I3 (8\% Mn, annealed, crosses); (c) calculated As $K$ edge x-ray absorption spectrum with Fermi energy set to $-$2.0~eV (full line) and $-$2.5~eV (points).}
\end{figure}

The As $K$ edge absorption and XMCD spectra, measured at a sample temperature of 7~K, are shown in Fig.~\ref{fig:2}. The absorption spectra are similar to those previously reported for GaAs, \cite{dalba93} with no significant differences in the near-edge part of spectra observed between the (Ga,Mn)As and (In,Ga,Mn)As samples. The XMCD manifests itself as a sharp Lorentzian-like peak which coincides with the onset of the As $K$ absorption edge. The sign of the XMCD and its lineshape are consistent with our previous observations. \cite{freeman08} However, its magnitude is strongly sample-dependent: in the most conductive sample (G4), it is around a factor of 3 larger than in our previous study, while in the insulating sample G2 it is scarcely visible.

The As $K$ edge absorption spectrum can be interpreted in terms of the part of the one-particle density of states that is accessible from the 1$s$ core level \cite{joly01,antonov10} (in contrast to shallower absorption edges, where multiplet effects are important \cite{vdl91}). Due to the electric dipole selection rules and the Pauli exclusion principle, the absorption spectrum is determined by the partial densities of unoccupied states with As 4$p$ character, and their corresponding transition probabilities. In semiconducting GaAs, the unoccupied 4$p$ states lie in the conduction band. However, if it is doped with acceptors, then this will introduce additional unoccupied states near $E_{\mathrm{F}}$, and transitions to these states will be allowed if they have As 4$p$ character. Hence, the presence of an XMCD signal at lower energy than the main absorption edge is an indicator of a density of unoccupied As 4$p$ states near the valence band edge in Mn-doped GaAs and (In,Ga)As. Furthermore, it also indicates that these states have a net orbital magnetic moment. Due to the absence of spin-orbit coupling in the core state, XMCD at the $K$ edge is sensitive to the orbital moment of the valence states, and not the spin moment. \cite{thole92,carra93}

To add support to the above picture, Fig.~\ref{fig:2}(c) shows As $K$ edge x-ray absorption spectra of GaAs calculated using the FDMNES code. \cite{joly01} The calculation was performed for a 99-atom cluster, using a finite-differences method to model the interatomic potentials. The x-ray absorption lineshape is qualitatively reproduced by the calculation. Most significantly, shifting the Fermi level in the calculation results in an increased intensity at the energy position of the measured XMCD peak. This strongly suggests that, even though the XMCD peak does not coincide with a distinct feature in the absorption spectrum, it is associated with transitions to states near $E_{\mathrm{F}}$ that become unoccupied due to $p$-type doping. 

More detailed first-principles calculations of the As $K$ edge absorption and XMCD in (Ga,Mn)As have recently been reported by Antonov \emph{et al.}. \cite{antonov10} The calculated XMCD spectrum essentially consists of a single sharp peak at the onset of the edge, in agreement with the present measurements. The XMCD was shown to arise due to a combination of the spin-orbit splitting of the As 4$p$ states and exchange splitting at the neighboring Mn site. \cite{antonov10}

\begin{figure}[h]    
\centering
\includegraphics[width=3.1in]{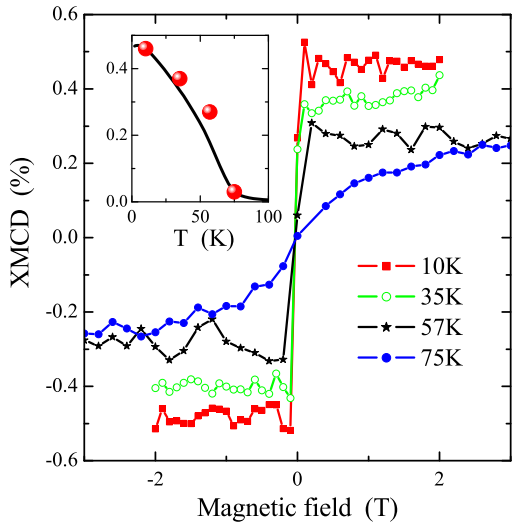}
\caption{\label{fig:3} (Color online)
XMCD at fixed x-ray energy of 11873~eV for sample I2 versus applied magnetic field, at a sample temperature of 10~K (squares), 35~K (circles), 57~K (stars) and 75~K (dots). The inset compares the temperature-dependence of the XMCD (points) and the magnetization measured by SQUID magnetometry (line) for a fixed field of 0.1~T, normalized at the lowest temperatures.}
\end{figure}

To confirm the dependence of the As XMCD on the ferromagnetic order of the Mn, we measured the dependence of the XMCD signal on the external magnetic field and temperature. The absorption signal was measured for left- and right-circular polarization, at a fixed energy $E$=11873eV corresponding to the XMCD peak. The difference as a function of magnetic field is shown in Fig.~\ref{fig:3}, for sample I2. A transition from a ferromagnetic to a paramagnetic response is observed on crossing $T_{\mathrm{C}}$. The inset compares the temperature dependence of the XMCD and the bulk magnetization obtained by SQUID magnetometry, at a fixed field of 0.1~T. The measurements are in agreement, showing that the As orbital moment is clearly associated with the magnetic order of the Mn ions.  

The integrated XMCD intensity can be related to the element-specific orbital magnetic moment by application of a sum rule. \cite{thole92} For the As $K$ edge, the 4$p$ orbital moment per As ion is given by $m_{\mathrm{orb,}4p}$ = $-(6-n_{4p})(2I_{\mathrm{XMCD}}/3I_{\mathrm{XAS}})$, where $I_{\mathrm{XMCD}}$ and $I_{\mathrm{XAS}}$ are the integrated XMCD signal over the absorption edge and the integrated absorption for 1$s$ to 4$p$ transitions respectively. $n_{4p}$ is the electron count in the 4$p$ shell, which we take to be equal to 3. We separate the 1$s$ to 4$p$ contributions to the absorption edge from the 1$s$ to continuum background by modelling the latter as a step function positioned under the highest intensity point of the absorption spectrum. The estimated 10-20\% uncertainties introduced by these assumptions are systematic, so that the relative values of the orbital moments per ion from the different samples can be compared with high accuracy.

\begin{figure}[h]    
\centering
\includegraphics[width=2.8in]{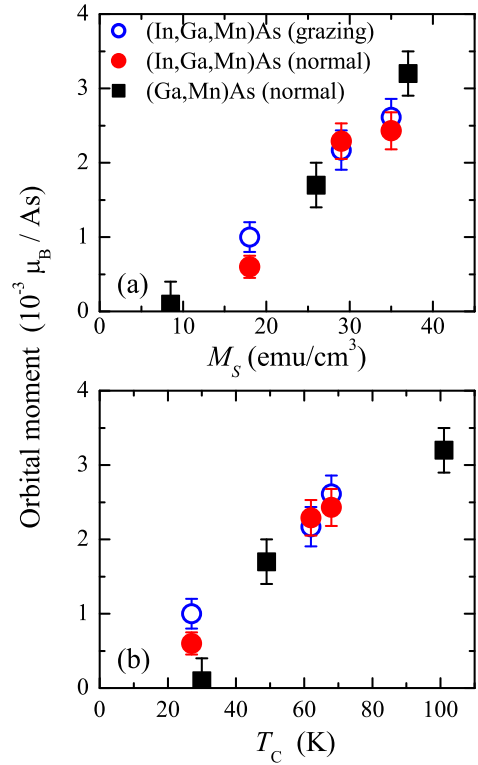}
\caption{\label{fig:4} (Color online)
Orbital magnetic moment per As ion versus (a) saturation magnetization $M_S$ and (b) Curie temperature $T_{\mathrm{C}}$ for a series of (Ga,Mn)As (squares) and (In,Ga,Mn)As (circles) samples. Filled symbols are for normal incidence and open symbols are for 15$^\circ$ incidence.}
\end{figure}

Figure~\ref{fig:4} shows the obtained orbital moments per As ion from sum rule analysis of the six samples studied. The result is plotted against the saturation magnetization $M_S$ [Fig.~\ref{fig:4}(a)] and the $T_{\mathrm{C}}$ [Fig.~\ref{fig:4}(b)] of the (Ga,Mn)As and (In,Ga,Mn)As films obtained by SQUID magnetometry. The As 4$p$ orbital moment increases approximately linearly with increasing $M_S$, which is  a measure of the concentration of Mn local magnetic moments that are participating in the ferromagnetism. This further indicates that the As polarization is induced by proximity to ferromagnetically ordered substitutional Mn. However, the As orbital moment falls to zero (within the experimental uncertainty) for finite values of $M_S$ and $T_{\mathrm{C}}$, suggesting that the presence of a significant As orbital polarization is not a necessary condition for ferromagnetism in (III,Mn)As materials.

Substitution of In for Ga results in competing effects which may influence the magnitude of the orbital moment on the As ion. On the one hand, it expands the lattice, leading to a weaker overlap of As and Mn orbitals. On the other hand, it enhances the valence band spin-orbit splitting, and also shifts the position of the host valence band edge relative to the Mn acceptor level. The similar magnitude of As 4$p$ orbital moment in (Ga,Mn)As and (In,Ga,Mn)As is likely to be due to the interplay of these competing effects.

\subsection{Mn $K$ edge XMCD}

\begin{figure}[h]    
\centering
\includegraphics[width=3.1in]{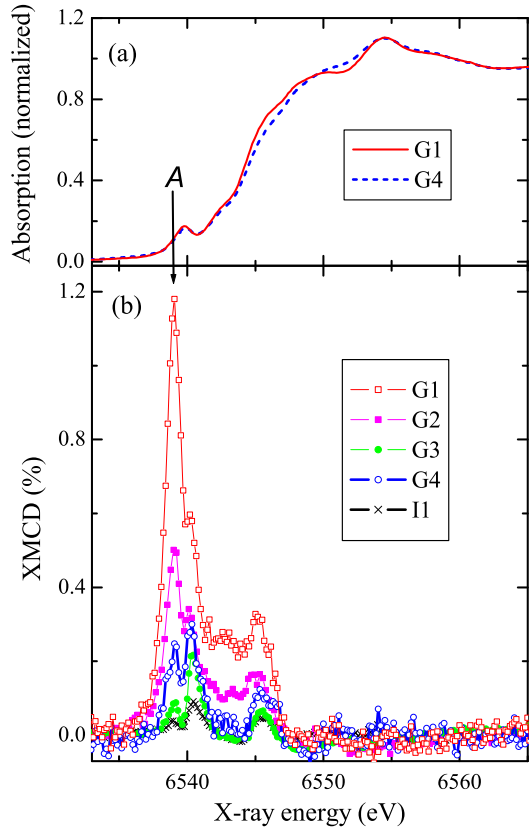}
\caption{\label{fig:5} (Color online)
(a) Mn $K$ edge x-ray absorption spectra, and (b) XMCD spectra, for (Ga,Mn)As samples G1 (1\% Mn, as-grown, open squares), G2 (2\% Mn, as-grown, filled squares), G3 (5\% Mn, as-grown, filled circles), G4 (5\% Mn, annealed, open circles), and (In,Ga,Mn)As sample I1 (8\% Mn, as-grown, crosses). The vertical arrow indicates the position of the Fermi edge peak $A$ referred to in the text.}
\end{figure}

The absorption and XMCD spectra recorded at the Mn $K$ edge at a sample temperature of 7~K are shown in Fig.~\ref{fig:5}. The absorption spectra are consistent with earlier studies, including the weak pre-edge peak centered at 6540~eV. \cite{titov05,bihler08} The largest contribution to the XMCD signal consists of a sharp double-peak structure centered on this pre-edge peak, together with a broader component of the same sign around 6545~eV. The most remarkable observation is that both the magnitude of the XMCD and the intensity ratio of the double-peak structure are strongly dependent on the Mn concentration. At low doping, where samples are on the insulating side of the metal-insulator transition, there is a dramatic increase in the intensity of the lowest energy XMCD peak, labelled $A$ in Fig.~\ref{fig:5}. At 1\% Mn doping the peak $A$ is around a factor of 5 larger than for the more metallic annealed 5\% film. The low-doped films also show increased intensity in the region between the second and third peaks in the XMCD spectra, centered around 6543~eV.

Pre-edge structures are a common feature in the $K$ edge absorption spectra of 3$d$ ions, and are usually associated with a 1$s$ to 3$d$ transition. While such a transition is formally electric-dipole forbidden, it can gain intensity through a mixing of 3$d$ and 4$p$ orbitals. \cite{westre97,goncharuk09} First principles calculations have shown that the pre-edge feature consists of two merged peaks, corresponding to transitions from the 1$s$ core level to two bands of hybridized Mn 4$p$ states. \cite{antonov10,goncharuk09} The lower energy peak corresponds to a spin-up band at $E_{\mathrm{F}}$, while the higher energy peak corresponds to an unoccupied spin-down band within the energy gap. The resulting XMCD spectrum is closely related to the energy-resolved Mn $4p$ orbital polarization per Mn ion. \cite{antonov10} The calculated XMCD spectrum, obtained in Ref.~\onlinecite{antonov10} using the local spin density approximation (LSDA), shows a comparable ratio of the pre-edge double peak heights to our measured result for low-doped films, with peak $A$ dominant in the spectrum.

In Ref.~\onlinecite{antonov10}, it was shown that the height of peak $A$ is considerably reduced if $E_{\mathrm{F}}$ is shifted towards the gap in a rigid-band picture. Therefore, the small size of the measured peak $A$ in the metallic (Ga,Mn)As films, reported previously by us, \cite{freeman08} was attributed to a Fermi level shift induced by compensating defects such as interstitial Mn. The increase of peak $A$ in the annealed film G4 compared to the as-grown film G3 is qualitatively consistent with this. However, this increase is small compared to the dramatic enhancement of peak $A$ in the insulating layers. It is known that the LSDA used in Ref.~\onlinecite{antonov10} tends to overestimate the 3$d$ weight of the states at $E_{\mathrm{F}}$. \cite{sandratskii04} In the metallic regime of (Ga,Mn)As, due to screening of the Coulomb interactions between the holes and acceptors, the states at $E_{\mathrm{F}}$  are not bound to the Mn ions, but are instead delocalized across the system.  

The above considerations suggest the following interpretation of the dependence of the Mn XMCD peak $A$ on doping. In the low doped, insulating regime, the valence holes are bound to the Mn acceptors, and are thus accessible from a Mn 1$s$ core level excitation. Below $T_{\mathrm{C}}$, these bound holes have a large orbital polarization, so that the XMCD at the onset of the Mn $K$ pre-edge is large. At higher doping, the valence holes are more extended, leading to a reduction in the probability of exciting a transition to these states from the 1$s$ core level. There is thus a clear connection between the XMCD signals at the lowest energies of the Mn and As $K$ edges: on crossing the metal-insulator transition, there is a transfer of the orbital polarization of the states near $E_{\mathrm{F}}$ from the Mn to the As ions; as a result, the XMCD at the Mn K edge decreases, and the XMCD at the As K edge increases.

It should be added that for sample I1, which also shows insulating behavior, the peak $A$ is very small. In this case, the Mn concentration is rather high, and the insulating behavior is due to a high degree of compensation. \cite{matsukura98} Therefore, the holes are bound to donor states that are not accessible from the Mn 1$s$ core level (e.g. the 4$s$ states of interstitial Mn \cite{yu02}), rather than to the substitutional Mn acceptors. 

\section{Conclusions}

In both the As and Mn $K$ edge XMCD spectra of (Ga,Mn)As and (In,Ga,Mn)As films, we observe features at the onset of the edge that are associated with transitions to states at or just above the Fermi level. The intensity of these features is a measure of the orbital magnetic moment per ion of the corresponding valence states. A marked dependence on the Mn doping is observed. In low doped samples, the Mn XMCD is large, and the As XMCD is small; with increasing doping, the lowest energy Mn XMCD peak decreases dramatically in intensity. The As XMCD increases with the saturation magnetization (which is proportional to the concentration of Mn that is participating in the ferromagnetic order) and the ferromagnetic Curie temperature.

These results are ascribed to a qualitative change in the character of the polarized states at $E_{\mathrm{F}}$ that occurs on crossing the transition from the low-doped insulating state to the high-doped metallic state in the (III,Mn)As systems. A transfer of the hole orbital magnetic moment from Mn to As sites may explain the success of 'host-like hole' models in calculating the magnetic and magnetotransport anisotropies, which are closely related to the valence band orbital polarization, in metallic (III,Mn)As materials. \cite{dietl01,rushforth07} Consistent with the present results, these models break down for low-doped insulating (Ga,Mn)As, where the holes have a hybridized Mn 4$p$ character. Anomalous properties of low-doped insulting films include giant anisotropic magnetoresistance and a perpendicular-to-plane magnetic anisotropy, \cite{wang05} and a qualitatively different dependence of $T_{\mathrm{C}}$ on the carrier density. \cite{sheu07} In spite of the very different conductivity and nature of the states at $E_{\mathrm{F}}$, ferromagnetism is observed in both regimes, with markedly lower $T_{\mathrm{C}}$ in the insulating case.

The XMCD at the Mn $K$ edge shows a clear signature of hole localization in the insulating regime of (Ga,Mn)As. It would be of interest to determine whether such behavior is observed in other transition metal doped semiconductors and insulators that are close to a metal-nonmetal transition. For example, alloying (Ga,Mn)As with AlAs shifts the transition to higher Mn concentrations due to the wider band gap of the semiconductor host, \cite{rushforth08} so that the large Mn $K$ edge XMCD may persist to higher doping. Similarly, (Ga,Mn)P shows hopping conductivity together with ferromagnetism up to the highest Mn concentrations studied; \cite{scarpulla05} for this material, due to hole localization the Mn $K$ edge XMCD may be expected to be much larger than for metallic (Ga,Mn)As, even though the Mn $L_{2,3}$ edge XMCD spectra (which is primarily sensitive to the Mn 3$d$ states) are very similar. \cite{stone06} The $K$ edge XMCD is a much more sensitive detector of hole localization than the unpolarized absorption spectrum, due to the sharply peaked orbital polarization at the Fermi level.

\begin{acknowledgments}
Funding from the UK EPSRC (EP/C526546/1), STFC (studentship grant CMPC07100), the Royal Society and the EU (grant NAMASTE-214499) is acknowledged.
\end{acknowledgments}

\end{document}